%% file: conference_101719.tex
\documentclass[conference]{IEEEtran}
\IEEEoverridecommandlockouts
\usepackage{cite}
\usepackage{amsmath,amssymb,amsfonts}
\usepackage{algorithm}
\usepackage{algorithmic}
\usepackage{graphicx}
\usepackage{textcomp}
\usepackage{color,bm}
\usepackage{array}
\usepackage{diagbox}
\def\BibTeX{{\rm B\kern-.05em{\sc i\kern-.025em b}\kern-.08em
    T\kern-.1667em\lower.7ex\hbox{E}\kern-.125emX}}

\definecolor{orange}{RGB}{255,107,0}

\input{sym_marco.tex}

\begin{document}

\title{Optimizing Fluid Antenna Configurations for Constructive Interference Precoding\\

\thanks{
The work was supported by the National Natural Science Foundation of China under Grant 62401340 and the Natural Science Foundation of Shandong Province under Grant ZR2023QF103. Corresponding authors: Mingjie Shao and Zhi Liu (email: mingjieshao@amss.ac.cn, liuzhi@sdu.edu.cn).}
}

\author{\IEEEauthorblockN{Wenxuan Sun\IEEEauthorrefmark{1}, Mingjie Shao\IEEEauthorrefmark{2}, Luteng Zhu\IEEEauthorrefmark{1},  Yao Ge\IEEEauthorrefmark{3}, Tong Zhang\IEEEauthorrefmark{4}, and Zhi Liu\IEEEauthorrefmark{1}
  \\
  \IEEEauthorblockA{\IEEEauthorrefmark{1} School of Information Science and Engineering, Shandong University, Qingdao, China\\
   \IEEEauthorrefmark{2} State Key Laboratory of Mathematical Sciences, AMSS, Chinese Academy of Sciences, Beijing, China\\
   \IEEEauthorrefmark{3} Continental-NTU Corporate Lab, Nanyang Technological University, Singapore\\
   \IEEEauthorrefmark{4} Guangdong Provincial Key Laboratory of Aerospace Communication and Networking Technology, HIT, Shenzhen, China
  }
  \\
  \vspace{-2mm}
  \IEEEauthorblockA{\footnotesize \{wenxuansun@mail.sdu.edu.cn, mingjieshao@amss.ac.cn, lutengzhu@mail.sdu.edu.cn, yao.ge@ntu.edu.sg, tongzhang@hit.edu.cn, liuzhi@sdu.edu.cn\} }
 }
}

\maketitle

\begin{abstract}
The fluid antenna system (FAS) has emerged as a new physical-layer concept to provide enhanced propagation conditions for multiuser multiple-input multiple-output (MIMO) communications over conventional 
fixed arrays. 
This work focuses on minimizing the maximum symbol error probability (SEP) under $M$-ary phase shift keying (MPSK) signaling in a multiuser downlink equipped with FAS, where each antenna moves within nonoverlapping intervals. 
This specific problem of joint SEP minimization with FAS and constructive interference (CI) precoding has not been previously addressed. 
The resulting problem turns out to be a nonconvex and nonsmooth optimization challenge. 
We transform the SEP minimization problem into a safety 
margin maximization problem in constructive interference precoding. 
Then, we customize a smoothing technique and a block coordinate descent (BCD) algorithm, with emphasis on low computational complexity. 
Simulation results show that our approach can reduce bit error rate (BER) compared to both the fixed arrays and FAS designed by existing particle swarm optimization (PSO). 
Also, our approach shows attractively low  computational complexity compared to PSO benchmarks.
\end{abstract}


\section{Introduction}

The relentless evolution of wireless communication systems imposes ever-increasing demands for higher data transmission rates, enhanced reliability, 
and ubiquitous connectivity. Meeting these demands necessitates significant improvements in spectral efficiency (SE), which has spurred considerable 
research efforts in recent years. One promising avenue for boosting SE involves actively improving the wireless propagation 
environment, thereby creating more favorable channel conditions for transmission. On one hand, the reconfigurable intelligent surfaces (RIS), 
which consist of large arrays of reconfigurable elements capable of dynamically shaping wireless signal propagation, have been extensively studied \cite{elmossallamy2020reconfigurable, shao2020minimum, pan2022overview}.
More recently, the concept of  fluid antenna system (FAS) has emerged as a novel paradigm offering a new degree of freedom for tailoring the communication channels \cite{hao2024fluid, wong2020fluid, new2024tutorial}.

FAS enhances quality of service (QoS) by adaptively adjusting antenna positions, with early studies theoretically demonstrating significant gains from this flexibility \cite{wong2023fluid}.  
Subsequent research has explored the application of FAS in various scenarios, demonstrating benefits for 
maximizing sum-rate \cite{cheng2024sum}, enabling 
integrated sensing and communication (ISAC)\cite{zhou2024fluid,zhang2024efficient}, and optimizing energy efficiency\cite{chen2023energy}. 
Please see an overview paper \cite{zhu2024historical } for more recent developments of the FAS.
However, optimizing antenna positions with minimum inter-antenna spacing poses significant challenges due to coupled constraints.
In the literature, there have been efforts based on successive 
convex approximation (SCA) using Taylor expansions \cite{hao2024fluid,zhou2024fluid}, heuristic algorithms like particle swarm optimization (PSO) \cite{kuang2024movable}, and deep reinforcement learning techniques \cite{wang2024fluid}. 
But these methods face the same issue of high computational complexity due to the mutual optimization of all the antenna positions. 

In addition to the advancements in antenna technologies, sophisticated precoding techniques are crucial for managing interference and maximizing signal quality 
in multiuser multiple-input multiple-output (MIMO) systems. 
Traditional linear precoding methods, such as zero-forcing (ZF)\cite{spencer2004zero} and weighted minimum mean-square error (WMMSE) \cite{zhao2023rethinking}, primarily focus on mitigating or eliminating inter-user interference. 
In contrast, the concept of constructive interference (CI) precoding emerged as a new precoding concept that exploits the possible benefits brought by interference on a symbol-by-symbol level \cite{masouros2009dynamic,jacobsson2017quantized,liu2022symbol}, which is later also termed as symbol-level precoding in the literature \cite{shao2019framework, sohrabi2018one, liu2022symbol, baghdady2002directional, kalantari2016directional}.
CI aims to manipulate interference, ensuring that signals combine constructively at each receiver from a symbol-detection perspective \cite{li2018interference,masouros2015exploiting}. 
Despite the individual promise shown by both FAS and CI precoding, the potential synergy gained by combining these two techniques remains unexplored.

This paper investigates this intersection: optimizing a multiuser multiple-input single-output (MISO) downlink 
system that leverages both FAS at the base station (BS) and CI precoding principle.
We adopt a simple, potentially more practically 
deployable fluid antenna model, where each antenna element adjusts its position 
within a predefined local interval in a linear array \cite{chen2023joint}. 
This approach seeks to strike a balance: it captures 
key benefits of antenna position adaptability while keeping the antenna position lightweight. 
Within this specific system configuration, we address 
the problem of minimizing the maximum symbol error probability (SEP) under $M$-ary phase shift keying (MPSK) modulation. 
The resulting problem turns out to be a nonconvex and nonsmooth optimization challenge. 
We transform the SEP minimization problem into a safety 
margin maximization problem by CI \cite{shao2018multiuser}. 
Then, we customize a smoothing technique and a block coordinate descent (BCD) algorithm, with emphasis on low computational overhead. 
Simulation experiments demonstrate that the proposed method 
significantly outperforms traditional CI method with fixed antenna array. 
Also, our proposed design achieves enhanced performance and significantly lower runtime complexity than existing fluid antenna position methods such as PSO.  
\section{System Model}
\begin{figure}[h]
	\centering
	\includegraphics[width=0.9\linewidth]{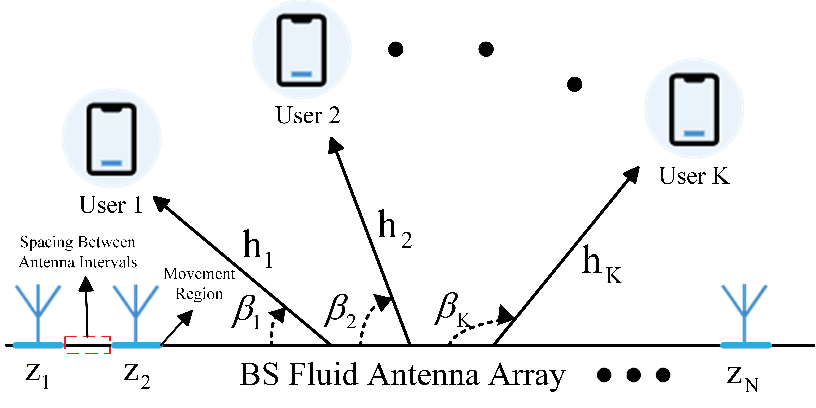}
	\caption{Illustration of system model}
	\label{fig:fluid-antenna}
\end{figure}

We consider a multiuser MISO downlink system, illustrated in Fig.\ref{fig:fluid-antenna}, where a base station (BS) with $N$ fluid antennas communicates with $K$ single-antenna users. 
FAS enables optimization of antenna positions to enhance QoS. Existing studies typically allow all antennas to move within a region with minimum inter-antenna spacing to maximize flexibility, but this increases optimization complexity due to coupled constraints.
On the other hand, it is also known that such a strategy leads to a coupling effect of the antenna positions and increases optimization difficulty. 
To balance QoS enhancement with optimization complexity, we propose optimizing each antenna within independent intervals. 
We consider a linear array and denote the position of the $n$th fluid antenna as $z_n$. 
The antenna position $z_n$ can be adjusted within the region 
\[
z_n \in \left[ \frac{(2n-1)D}{2N}-\delta\frac{D}{2N}, \frac{(2n-1)D}{2N}+\delta\frac{D}{2N} \right],
\]
where $[0,D]$ is the whole linear region, {and $\frac{(1-\delta) D}{N}$ is the spacing between successive antenna intervals, which is illustrated in Fig.~\ref{fig:fluid-antenna}.}
The assembled antenna position vector is defined as $\bm z = (z_1, z_2, \cdots, z_N )^{\top} \in \mathbb{R}^N$.

We consider a far-field communication scenario. The downlink channel   from the BS to the $i$th user follows a line-of-sight (LOS) model, which can be modeled as 
\begin{equation}
	\boldsymbol{h}_i = \alpha_i \boldsymbol{a}(\boldsymbol{z}, \beta_i),\quad i = 1,\dots,K,
\end{equation}
where $\alpha_i$ denotes the propagation gain and $\beta_i \in [0,\pi]$ is the departure angle for the $i$th user. The steering vector 
$\boldsymbol{a}(\boldsymbol{z}, \beta_i) \in \mathbb{C}^N$ is defined by
\begin{equation}\label{eq:steer}
	\boldsymbol{a}(\boldsymbol{z}, \beta_i) = \begin{bmatrix}
		e^{-j \frac{2\pi}{\lambda} \cos(\beta_i) z_1} \\
		e^{-j \frac{2\pi}{\lambda} \cos(\beta_i) z_2} \\
		\vdots \\
		e^{-j \frac{2\pi}{\lambda} \cos(\beta_i) z_N}
	\end{bmatrix},
\end{equation}
where $\lambda$ is the signal wavelength. 

The received signal at the $i$th user during symbol time $t$ is expressed as
\begin{equation}
	y_{i,t} = \boldsymbol{h}_i^\top\boldsymbol{x}_t + \eta_{i,t}\quad i=1,\cdots,K,\ t=1,\cdots,T,
\end{equation}
where $\boldsymbol{x}_t \in \mathbb{C}^N$ is the transmitted signal vector from the BS, $y_{i,t} \in \mathbb{C}$ is the received signal, 
and $\eta_{i,t} \sim \mathcal{CN}(0, \sigma^2)$ denotes circularly symmetric 
complex Gaussian noise. Here, $T$ 
represents the length of the transmission block.

For the user signals, we adopt MPSK modulation, where the transmitted 
symbol $s_{i,t}$ belongs to the constellation set
\begin{equation}
	s_{i,t} \in \mathcal{S} \triangleq \left\{ s \mid s = e^{j \frac{2\pi m }{M}}, \; m = 0, \ldots, M-1 \right\}.
\end{equation}
Let dec: $\mathbb{C} \rightarrow \mathcal{S}$ denote the MPSK decision function, i.e. 
$\text{dec}(y) = e^{j\frac{2\pi m}{M}}$, where $m \in \{0, \ldots, M-1\}$ is such that the phase angle of
$y$, lies in $\left[\frac{2m\pi}{M} - \frac{\pi}{M}, \frac{2m\pi}{M} + \frac{\pi}{M}\right]$. 
At the users' side, each user detects its symbol by $\hat{s}_{i,t} = \text{dec}(y_{i,t})$.

In this paper, we aim to jointly optimize the precoder from user symbols $\{s_{i,t}\}$ to the transmitted signal $\{\bx_t\}$ and the antenna position $\bz$  such that the SEPs at all the users are minimized.

\section{Problem Formulation}

The SEP for the $i$th user at 
time $t$ is defined as 
\begin{equation}\label{eq:SEP}
	\text{SEP}_{i,t} = \Pr(\hat{s}_{i,t} \neq s_{i,t} | s_{i,t}),
\end{equation}
where $\hat{s}_{i,t}$ is the symbol decoded by the $i$th user, and $s_{i,t}$ is the transmitted symbol. 
We aim to minimize the worst-case SEP under a transmit 
power and antenna position constraint, formulating the optimization problem as
\begin{equation}\label{eq:SEPP}
	\begin{aligned}
		& \min_{\boldsymbol{X}, \boldsymbol{z}} \max_{\substack{t=1,\ldots,T\\i=1,\ldots,K}} \text{SEP}_{i,t} \\
		& \text{s.t. } \|\boldsymbol{x}_t\|^2 \leq P, \forall t \in \mathcal{T}, \boldsymbol{z} \in \mathcal{Z},
	\end{aligned}
\end{equation}
where 
\begin{equation*}
  \begin{split}
    \mathcal{Z} = & \left\{\bz\mid z_n 
\in \left[ \frac{(2n-1)D}{2N}-\delta\frac{D}{2N}, \frac{(2n-1)D}{2N}+\delta\frac{D}{2N} 
\right] \right\} , 
  \end{split}
\end{equation*}
$P$ is the maximum allowable transmit power, $\boldsymbol{X}=\{\boldsymbol{x}_t\}_{t=1}^T$ and   $\mathcal{T}=  \{1,2,\ldots,T\}$.

The SEP \eqref{eq:SEP} for MPSK signaling does not admit an explicit form. 
We resort to an effective approximation to the SEP, which is shown as \cite{shao2018multiuser, wu2023diversity}
\begin{equation}\label{eq:SEP_bound}
	Q \left( \frac{\sqrt{2} \sin \frac{\pi}{M} \alpha_{i,t}}{\sigma} \right) \leq \text{SEP} \leq 2 Q \left( \frac{\sqrt{2} \sin \frac{\pi}{M} \alpha_{i,t}}{\sigma} \right),
\end{equation}
where $Q(x) = \frac{1}{\sqrt{2\pi}} \int_x^\infty e^{-\frac{1}{2}x^2} dx$, $\alpha_{i,t}$ is called the {\it safety margin} and is defined as 
$$
\alpha_{i,t} = \Re \{ \boldsymbol{h}(\boldsymbol{z})_i^\top \boldsymbol{x}_t s_{i,t}^*
\} - | \Im \{ \boldsymbol{h}(\boldsymbol{z})_i^\top \boldsymbol{x}_t s_{i,t}^* \} | \cot \left( \frac{\pi}{M} \right).
$$
Here, $\Re\{\cdot\}$ and $\Im\{\cdot\}$ denote the real and imaginary parts, 
respectively, and $s_{i,t}^*$ is the complex conjugate of $s_{i,t}$. 

As a result, instead of directly minimizing the SEP, we seek to maximize the safety margin, which leads to the following design problem 
\begin{equation}\label{eq:SAFETYP}
	\begin{aligned}
		&\max_{\boldsymbol{X},\boldsymbol{z}} \min_{\substack{t=1,\ldots,T\\i=1,\ldots,K}} \alpha_{i,t} \\
		&\text{s.t.} \ ||\boldsymbol{x}_t||^2 \leq P, \forall t \in \mathcal{T}, \boldsymbol{z}\in\mathcal{Z},
	\end{aligned}
\end{equation}

To facilitate the expression, we reformulate \eqref{eq:SAFETYP} by converting 
the complex vector $\boldsymbol{x}_t$ into a real-valued vector 
$\bar{\boldsymbol{x}}_t = [\Re \{ \boldsymbol{x}_t \}^\top, \Im \{ \boldsymbol{x}_t \}^\top]^\top$. 
Define auxiliary vectors
\begin{equation}\label{eq:LP}
  \begin{split}
    &\boldsymbol{b}(\boldsymbol{z})_{i,t} = [\Re \{ s_{i,t}^* \boldsymbol{h}(\boldsymbol{z})_i^\top \}, -\Im \{ s_{i,t}^* \boldsymbol{h}(\boldsymbol{z})_i^\top \}]^\top,\\
	&\boldsymbol{r}(\boldsymbol{z})_{i,t} = \cot \left( \frac{\pi}{M} \right) [\Im \{ s_{i,t}^* \boldsymbol{h}(\boldsymbol{z})_i^\top \}, \Re \{ s_{i,t}^* \boldsymbol{h}(\boldsymbol{z})_i^\top \}]^\top,\\
	&\boldsymbol{u}(\boldsymbol{z})_{i,t} = -\boldsymbol{b}(\boldsymbol{z})_{i,t} +\boldsymbol{r}(\boldsymbol{z})_{i,t}, 
	\boldsymbol{w}(\boldsymbol{z})_{i,t} = -\boldsymbol{b}(\boldsymbol{z})_{i,t} - \boldsymbol{r}(\boldsymbol{z})_{i,t},
  \end{split}
\end{equation}
where $\boldsymbol{b}(\boldsymbol{z})_{i,t}, \boldsymbol{r}(\boldsymbol{z})_{i,t}, \boldsymbol{u}(\boldsymbol{z})_{i,t}, \boldsymbol{w}(\boldsymbol{z})_{i,t} 
\in \mathbb{R}^{2N}$. Then the problem \eqref{eq:SAFETYP} can be rewritten as
\begin{equation}\label{eq:RP1}
	\begin{aligned}
		&\min_{\bar{\boldsymbol{X}},\boldsymbol{z}}\max_{\substack{t=1,\ldots,T\\i=1,\ldots,K}}
		\max \left\{ \boldsymbol{u}(\boldsymbol{z})_{i,t}^\top \bar{\boldsymbol{x}}_t, 
		\boldsymbol{w}(\boldsymbol{z})_{i,t}^\top \bar{\boldsymbol{x}}_t \right\} \\
		&\text{s.t.} \ ||\bar{\boldsymbol{x}}_t||^2 \leq P, \forall t \in \mathcal{T}, \boldsymbol{z}\in\mathcal{Z},
	\end{aligned}
\end{equation}
where $\bar{\boldsymbol{X}}=\{\bar{\boldsymbol{x}}_t\}_{t=1}^T$.  For each $t$, we define $\boldsymbol{v}_{j,t} 
\in \mathbb{R}^{2N}$ for $j = 1, \ldots, 2K$ as
\begin{equation}
	\boldsymbol{v}(\boldsymbol{z})_{j,t} = 
	\begin{cases}
		\boldsymbol{u}(\boldsymbol{z})_{j,t} \quad j=1, 3, 5, \cdots, 2K-1,\\
		\boldsymbol{w}(\boldsymbol{z})_{j,t} \quad j=2, 4, 6, \cdots, 2K.
	\end{cases}
\end{equation}
Problem (9) can be transformed into a minmax problem as
\begin{equation}\label{eq:nonsmooth}
	\begin{aligned}
		&\min_{\bar{\boldsymbol{X}},\boldsymbol{z}} \max_{\substack{t=1,\ldots,T\\j=1,\ldots,2K}} 
		\boldsymbol{v}(\boldsymbol{z})_{j,t}^\top 
		\bar{\boldsymbol{x}}_t\\
		&\text{s.t.} \ \| \bar{\boldsymbol{x}}_t \|^2 \leq P, \forall t \in \mathcal{T}, \boldsymbol{z}\in\mathcal{Z} .
	\end{aligned}
\end{equation}
The problem \eqref{eq:nonsmooth} has a nonsmooth objective function due to the $\max$ function.
Also, the optimization of antenna position $\bz$ is implicitly embedded in the $\bv(\bz)$ in a nonlinear manner by the model \eqref{eq:steer} and \eqref{eq:LP}.
These factors make the design problem \eqref{eq:nonsmooth} challenging to solve. 

\section{Reformulation and Algorithm Design}
\subsection{Smooth Approximation}

We adopt the smooth technique in \cite{nesterov2005smooth} to smooth the objective function in \eqref{eq:nonsmooth}, which will allow us to leverage the optimization techniques in smooth optimization. 
The smoothing approximation of problem \eqref{eq:nonsmooth} is given by 
\begin{equation}\label{eq:finalproblem}
	\begin{aligned}
		&\min_{\bar{\boldsymbol{X}},\boldsymbol{z}} \psi(\bar{\boldsymbol{X}},\boldsymbol{z})\\
		&\text{s.t.} \ \| \bar{\boldsymbol{x}}_t \|^2 \leq P, \forall t \in \mathcal{T}, \boldsymbol{z}\in\mathcal{Z},
	\end{aligned}
\end{equation}
where
\begin{equation*}
  \begin{split}
    \psi(\bar{\boldsymbol{X}}, \boldsymbol{z}) = &~\max_{\boldsymbol{\lambda} \in \Delta_{2K T}} 
\sum_{t \in \mathcal{T}} \sum_{j \in \mathcal{J}}\left(\lambda_{j,t} 
\boldsymbol{v}(\boldsymbol{z})_{j,t}^\top \bar{\boldsymbol{x}}_t - \frac{\mu}{2} \lambda_{j,t}^2 
\right), \\
\Delta_{2K T} = &~\left\{ \boldsymbol{\lambda} \in \mathbb{R}^{ 2K T} \mid 
\sum_{t \in \mathcal{T}} \sum_{j \in \mathcal{J}}  \lambda_{j,t} = 1, 
\lambda_{j,t} \geq 0 \right\},
  \end{split}
\end{equation*}
where $\boldsymbol{\lambda} = \{\lambda_{j,t}\}_{j\in \mathcal{J}, t \in \mathcal{T}}$ is a weight  vector belonging to the $2KT$-dimensional simplex $\Delta_{2KT}$, $\mu > 0$ is a small positive parameter used 
to control the degree of smoothing. 
When it approaches 0, the function $\psi(\bar{\boldsymbol{X}},\boldsymbol{z})$ gradually approximates the original objective function. 

With the smoothing technique, the  gradients of the objective function with respect to $\bar{\boldsymbol{X }}$ and  $\boldsymbol{z}$ can be computed by
\begin{equation}\label{eq:precodinggradient}
	\nabla_{\bar{\boldsymbol{x}}_t} \psi(\bar{\boldsymbol{X}},\boldsymbol{z}) = 
	\sum_{j\in\mathcal{J}} \lambda_{j,t}^*
	\boldsymbol{v}(\boldsymbol{z})_{j,t}, ~ \forall t \in \mathcal{T},
\end{equation}
\begin{equation}\label{eq:positiongradient}
	\nabla_{\boldsymbol{z}} \psi(\bar{\boldsymbol{X}}, \boldsymbol{z}) = 
	\sum_{t \in \mathcal{T}} \sum_{j \in \mathcal{J}} \lambda_{j,t}^*(\bar{\boldsymbol{x}}_t^\top\frac{\partial \boldsymbol{v}(\boldsymbol{z})_{j,t}}{\partial \boldsymbol{z}}),
\end{equation}
where $\boldsymbol{\lambda}^*$ is the unique optimal 
solution of the inner problem, $\mathcal{J}=\{1,2,\ldots,2K\}$. The problem to get 
$\boldsymbol{\lambda}^*$
can be expressed as
\begin{equation}
	\boldsymbol{\lambda}^* = \arg\max_{\boldsymbol{\lambda} \in \Delta_{2KT}}  \sum_{t \in \mathcal{T}} \sum_{j \in \mathcal{J}}\left( \lambda_{j,t} 
	\boldsymbol{v}(\boldsymbol{z})_{j,t}^\top \bar{\boldsymbol{x}}_t - \frac{\mu}{2}  \lambda_{j,t}^2 \right).
\end{equation}
By checking  the Karush-Kuhn-Tucker (KKT) conditions, we can obtain the optimal $\bm \lambda^*$ in closed form
\begin{equation}\label{eq:closeform}
	\lambda_{j,t}^* = \max\Big\{ \frac{\boldsymbol{v}(\boldsymbol{z})_{j,t}^\top \bar{\boldsymbol{x}}_t + \gamma}{\mu} , 0\Big\},
\end{equation}
where  $\gamma$ is the Lagrange multiplier in the Lagrangian function. Its value can be determined by using the bisection method.

\subsection{Joint Precoding and Antenna Position Design Algorithm}

Jointly optimizing $\bar{\boldsymbol{X}}$ and $\boldsymbol{z}$ is challenging due to the coupling effect in the objective function. 
To address this, we employ the BCD method, which alternates between optimizing  $\bar{\boldsymbol{X}}$ and $\boldsymbol{z}$ while holding the other variable fixed. 
The BCD method is shown in Algorithm~\ref{alg:bcd}.

\begin{algorithm}[htb!]
	\caption{The BCD Method for Problem \eqref{eq:finalproblem}}
	\label{alg:bcd}
	\begin{algorithmic}[1]
		\STATE Initialize: \(\bar{\boldsymbol{X}}_0 = \{\bar{\boldsymbol{x}}_{t,0}\}_{t \in \mathcal{T}}\), \(\boldsymbol{z}_0\), convergence threshold \(\epsilon\), iteration index \(\ell = 0\).
		\REPEAT
		\STATE update \(\ell \gets \ell+1\).
		\STATE \(\text{Block 1: Update precoding vectors} \ \bar{\boldsymbol{X}}_{\ell+1}
		\newline \bar{\boldsymbol{X}}_{\ell+1} \gets \arg\min_{\bar{\boldsymbol{X}}} 
		\psi(\bar{\boldsymbol{X}}, \boldsymbol{z}_{\ell}) \newline \hspace*{4em} \text{s.t.} \ \| \bar{\boldsymbol{x}}_t \|^2 \leq P, \forall t \in \mathcal{T}\).
		\STATE \(\text{Block 2: Update antenna positions} \ \boldsymbol{z}_{\ell+1} \newline 
		\boldsymbol{z}_{\ell+1} \gets \arg\min_{\boldsymbol{z}} 
		\psi(\bar{\boldsymbol{X}}_{\ell+1}, \boldsymbol{z}) \quad 
		\text{s.t.} \ \boldsymbol{z} \in \mathcal{Z}\).
		\UNTIL \(\max_{t} \| \bar{\boldsymbol{x}}_{t, \ell+1} - \bar{\boldsymbol{x}}_{t, \ell} \| \leq \epsilon\) and \(\| \boldsymbol{z}_{\ell+1} - \boldsymbol{z}_{\ell} \| \leq \epsilon\).
	\end{algorithmic}
\end{algorithm}

The BCD algorithm decomposes the joint optimization into two subproblems which are 
solved iteratively.
In the following, we specify the solving algorithms for each subproblem. 
 In the $(\ell+1)$th iteration, given the current antenna position vector 
$\boldsymbol{z}_{\ell}$, we  update the  precoding vectors by solving
\begin{equation}\label{eq:precoding}
	\begin{aligned}
		 \bar{\boldsymbol{X}}_{\ell+1}=&~\arg\min_{\bar{\boldsymbol{X}}} \psi
		(\bar{\boldsymbol{X}}, \boldsymbol{z}_{\ell}) \\
		 \text{s.t.} &~ \| \bar{\boldsymbol{x}}_t \|^2 \leq P, \forall t \in \mathcal{T}.
	\end{aligned}
\end{equation}

To solve the precoding subproblem efficiently, we implement the accelerated projected gradient (APG) method. 
It is noted that the precoding problem can be decoupled across the time $t$ since both the objective function and constraints can be decoupled.
For each $\bar{\bx}_t$, the APG method for solving precoding subproblem is
\begin{equation}
	\bar{\boldsymbol{x}}_{t,k+1} = \Pi_{\mathcal{X}} \left( \boldsymbol{y}_{t,k} - {\kappa_{t,k}} \nabla_{\bar{\bx}_t}\psi(\bar{\boldsymbol{X}}_{k},\boldsymbol{z}_{\ell}) \right),
\end{equation}
where $\kappa_t$ represents the step size, $\Pi_\mathcal{X}$ denotes the projection of each \(\bar{\boldsymbol{x}}_t\) onto 
\(\{ \bar{\boldsymbol{x}}_t \mid \|\bar{\boldsymbol{x}}_t\|^2 \leq P \}\), and $\boldsymbol{y}_{t,k}$ is an extrapolated point and is given by
\begin{equation}
	\boldsymbol{y}_{t,k} = \bar{\boldsymbol{x}}_{t,k} + s_k (\bar{\boldsymbol{x}}_{t,k} - \bar{\boldsymbol{x}}_{t,k-1}),
\end{equation}
with
\begin{equation}\label{eq:stau}
	s_k = \frac{\tau_{k-1} - 1}{\tau_k}, \quad \tau_k = \frac{1 + \sqrt{1 + 4\tau_{k-1}^2}}{2}.
\end{equation}
It can be derived from \cite{nesterov2005smooth}  that the Lipschitz constant of the gradient $\nabla_{\bar{\bx}_t}\psi(\bar{\boldsymbol{X}}_k,\boldsymbol{z}_{\ell})$ is given by  
$L_t = \frac{1}{\mu} \left\| \boldsymbol{V}_t \right\|_2^2,
$
where $\boldsymbol{V}_t = [\boldsymbol{v}(\boldsymbol{z})_{1,t}, \ldots, \boldsymbol{v}(\boldsymbol{z})_{2K,t}]$. 
We choose $\kappa_t=\frac{1}{L_t}$ to guarantee global convergence.
The  APG method for updating the precoding vectors is summarized in Algorithm \ref{alg:APG}.

\begin{algorithm}[h]
	\caption{The APG Method for Problem \eqref{eq:precoding}}
	\label{alg:APG}
	\begin{algorithmic}[1]
	\STATE \textbf{Input}: Initial \(\bar{\boldsymbol{X}}_0 = \{\bar{\boldsymbol{x}}_{t,0}\}_{t \in \mathcal{T}}\), \(\bar{\boldsymbol{X}}_{-1} = \bar{\boldsymbol{X}}_0\),  
	\(\tau_0 = 1\), fixed position vector \(\boldsymbol{z}_{\ell}\),  iteration index \(k=0\).
	\REPEAT
	\STATE Update \(k \gets k + 1\).
	\STATE Compute \(\tau_k, s_k\) according to \eqref{eq:stau}.
	\FORALL{\( t \in \mathcal{T} \)}
		\STATE Compute extrapolation point \newline\(\boldsymbol{y}_{t,k} = \bar{\boldsymbol{x}}_{t,k} + s_k (\bar{\boldsymbol{x}}_{t,k} - \bar{\boldsymbol{x}}_{t,k-1})\).
		\STATE Compute gradient \(\nabla_{\bar{\boldsymbol{x}}_t} \psi(\bar{\boldsymbol{X}}_k, \boldsymbol{z}_{\ell})\) via \eqref{eq:precodinggradient}.
		\STATE Update \(\bar{\boldsymbol{x}}_{t,k+1} = \Pi_{\mathcal{X}} \left( \boldsymbol{y}_{t,k} - \kappa_t \nabla_{\bar{\boldsymbol{x}}_t} \psi(\bar{\boldsymbol{X}}_{k},\boldsymbol{z}_{\ell}) \right)\).
	\ENDFOR
	\UNTIL \(\max_{t} \| \bar{\boldsymbol{x}}_{t,k+1} - \bar{\boldsymbol{x}}_{t,k} \| \leq \epsilon\).
	\end{algorithmic}
\end{algorithm}

With the updated precoding vectors $\bar{\boldsymbol{X}}_{\ell+1}$, we solve the following subproblem to update the antenna positions
\begin{equation}\label{eq:position}
	\boldsymbol{z}_{\ell+1}=\arg\min_{\boldsymbol{z}} \psi(\bar{\boldsymbol{X}}_{\ell+1}, 
	\boldsymbol{z}), \quad \text{s.t.} \quad \boldsymbol{z} \in \mathcal{Z}.
\end{equation}

For this subproblem, we employ a projected gradient descent (PGD) method with backtracking line search \cite{boyd2004convex}. 
The PGD method iteratively minimizes the objective function \(\psi(\bar{\boldsymbol{X}}_{\ell+1}, \boldsymbol{z})\) while ensuring 
\(\boldsymbol{z} \in \mathcal{Z}\) for the antenna position subproblem. At each step, it computes the gradient \(\nabla_{\boldsymbol{z}} \psi\), updates along 
the negative gradient direction, and projects the result onto the feasible set \(\mathcal{Z}\) via element-wise thresholding to satisfy the disjoint interval 
constraints for antenna positions, continuing until the change in \(\boldsymbol{z}\) is sufficiently small.

Since we assume the antennas lie in disjoint intervals,  the computation of the projection $\Pi_{\cal Z}$ can be done element-wise by thresholding, which can be expressed as 
\begin{equation*}
  \hat{\bz} = \Pi_{\cal Z}(\bz) \Leftrightarrow \hat{z}_n =  [z_n]_{l_b}^{u_b},
\end{equation*}
where $l_b=\frac{(2n-1)D}{2N}-\delta\frac{D}{2N}$, $u_b=\frac{(2n-1)D}{2N}+\delta\frac{D}{2N}$, the notation $[z]_{a}^b$ denotes the thresholding operation $[z]_{a}^b = \max\{ \min\{ z,b  \}, a\}$. 
This alleviates the computational overhead compared to jointly optimizing all the antenna positions in a shared region. 
The algorithm for updating antenna positions is summarized 
in Algorithm \ref{alg:GD}.

\begin{algorithm}[H]
	\caption{The PGD Method for Problem \eqref{eq:position}}
	\label{alg:GD}
	\begin{algorithmic}[1]
		\STATE Initialize \(\boldsymbol{z}_0\), fixed transmitted signal matrix $\bar{\boldsymbol{X}}_{\ell+1}$, convergence threshold \(\epsilon\), iteration index \(k=0\).
		\REPEAT
		\STATE update \(k \gets k+1\).
		\STATE compute gradient \(\nabla_{\boldsymbol{z}}\psi(\bar{\boldsymbol{X}}_{\ell+1} , 
		\boldsymbol{z}) \) according to \eqref{eq:positiongradient}.
		\STATE Apply backtracking line search to find step length \(\eta_B\)
		\STATE  update \(\boldsymbol{z}_{k+1} = \Pi_\mathcal{Z}(\boldsymbol{z}_{k} - \eta_B \nabla_{\boldsymbol{z}} \psi(\bar{\boldsymbol{X}}_{\ell+1}))\).
		\UNTIL \(\| \boldsymbol{z}_{k+1} - \boldsymbol{z}_{k} \| \leq \epsilon\).
	\end{algorithmic}
\end{algorithm}

\subsection{Computational Complexity}

We analyze the computational complexity of the proposed algorithms. Table \ref{table:I} summarizes the per-iteration complexities of the APG method for Problem \eqref{eq:precoding} and the PGD method for problem \eqref{eq:position}. It is worth noting that the computations of the gradient and projection operations contribute to the main complexity.
\begin{table}[t]
 \centering
 \caption{computational complexity of our proposed algorithm}\label{table:I}
 \resizebox{\linewidth}{!}{%
  \begin{tabular}{>{\centering\arraybackslash}m{18mm}|>{\centering\arraybackslash}m{15mm}|>{\centering\arraybackslash}m{15mm}|>{\centering\arraybackslash}m{15mm}}
   \hline
     Subproblem and Method & Gradient Calculation & Projection Calculation & \ Per-iteration Complexity\\ \hline\hline
     APG for \eqref{eq:precoding}&  ${\cal O}(KNT)$ &  ${\cal O}(NT)$  &  ${\cal O}(KNT)$\\ \hline
    PGD for \eqref{eq:position}&  ${\cal O}(KN^2T)$  &  ${\cal O}(N)$ &  ${\cal O}(KN^2T)$ \\ \hline\hline
 \end{tabular}}
\end{table}

\section{Simulation Results}

In this section, we evaluate the performance of our proposed joint constructive interference precoding and antenna position (CIAP) optimization algorithm through numerical simulations. We compare our proposed CIAP method with two baseline schemes: (1) a fixed position array, where the BS antennas are uniformly distributed on a linear array and the CI precoding is optimized using the same algorithm as detailed in Algorithm~\ref{alg:APG}, namely ``FPA" in the legend;
(2) a FA scheme where the antenna position is optimized by the state-of-the-art PSO approach and the CI precoding is optimized by the APG Algorithm \ref{alg:APG}, namely ``BCD-PSO".

The simulation parameters are set as follows: The total length of 
the antenna deployment line is $D = \lambda N$, where the wavelength $\lambda=0.01$m \cite{zhu2023modeling}.
In this setting, the parameter $\delta \in [0,0.5]$ is set to satisfy the minimum spacing requirement $\frac{\lambda}{2}$ between antennas. In our simulations,  the default position adjustment factor $\delta=0.1$ unless 
stated otherwise.
The departure angles $\beta_i$ for the $K$ users are drawn independently and uniformly from the range $[0,\pi]$. 
The transmission block length is $T=5$. 
The transmit power constraint is $P=1$W. The smoothing parameter 
$\mu = 0.3+log(1+\sigma)$. 
We evaluate the BER performance over $500$ random Monte-Carlo trials.
The signal-to-noise ratio is defined as 
$
\text{SNR} = \frac{P}{\sigma^2}.
$

\begin{figure}[t]
	\centering
	\includegraphics[width=0.9\linewidth]{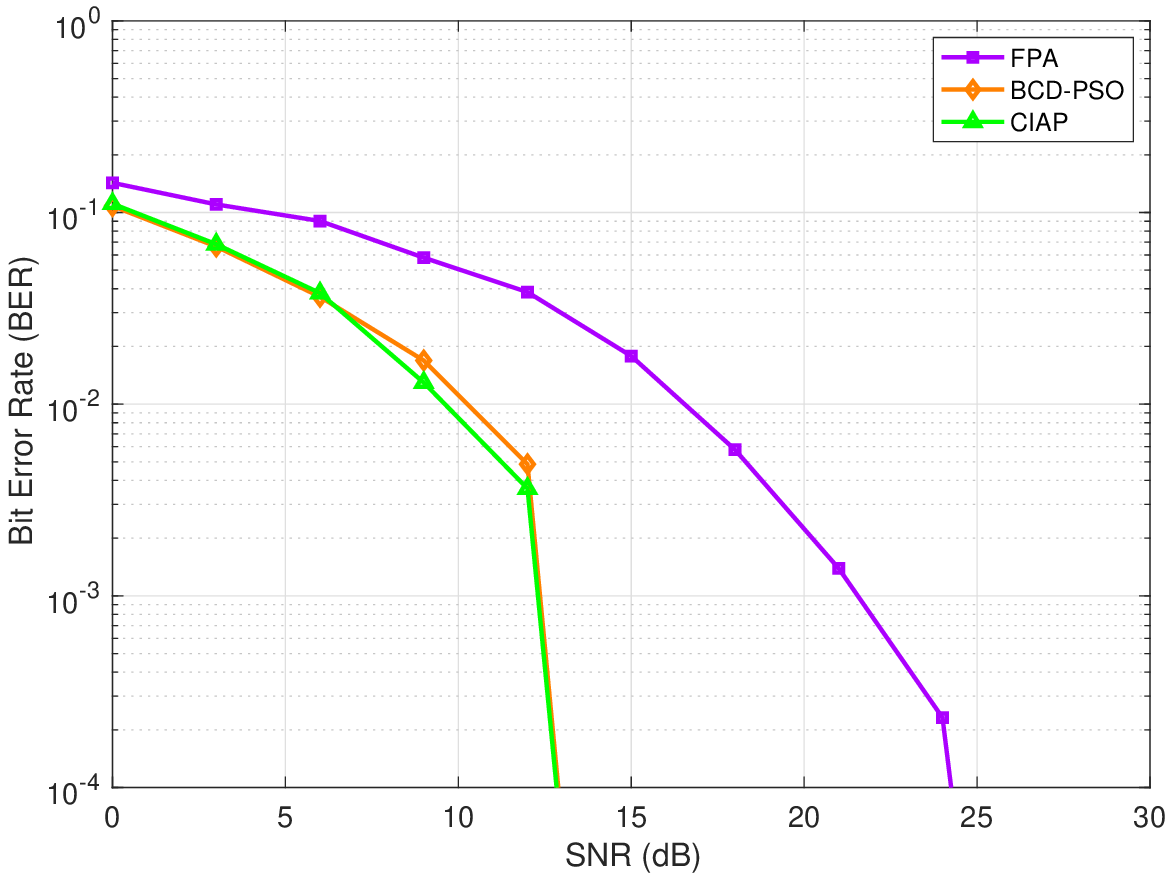}
	\caption{Average BER performance versus SNR; QPSK}
	\label{fig:10_10_01_5_4PSK}
\end{figure}
\begin{figure}[t]
	\centering
	\includegraphics[width=0.9\linewidth]{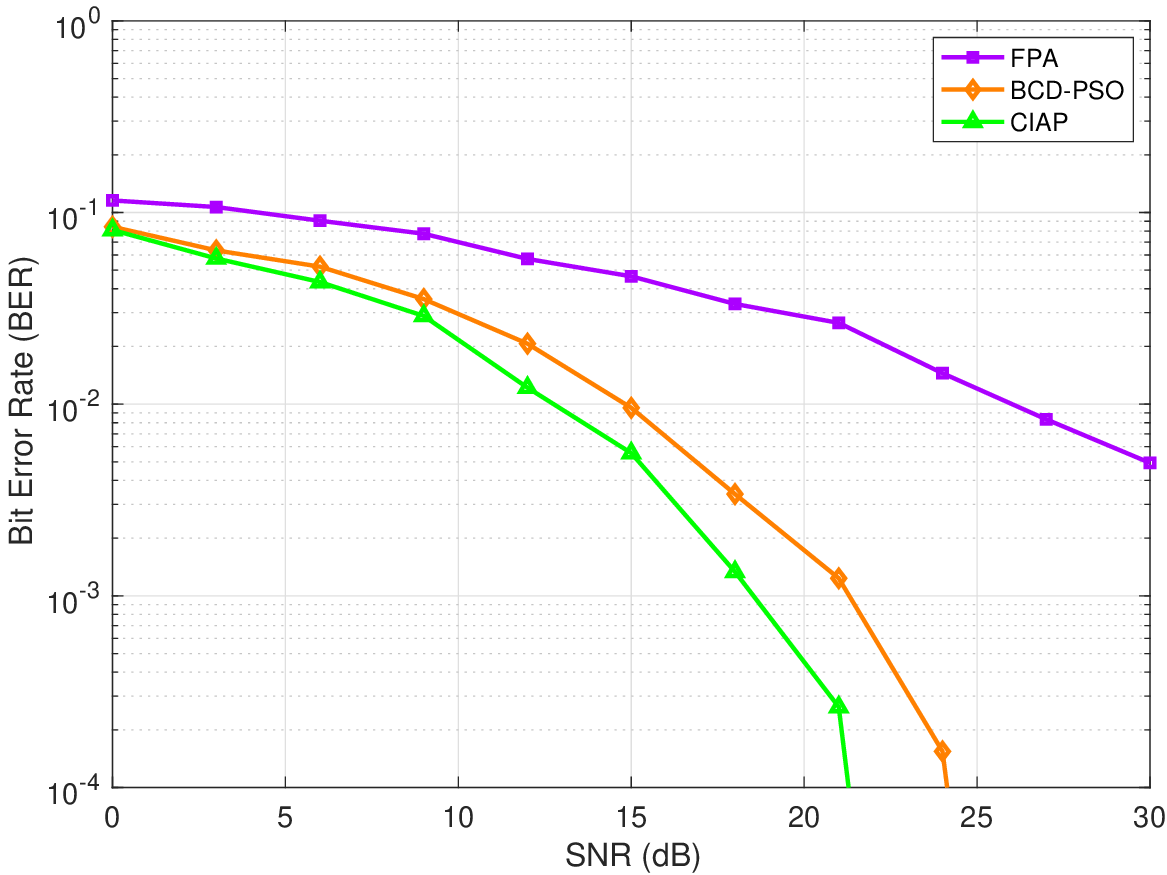}
	\caption{Average BER performance versus SNR; 8PSK}
	\label{fig:10_10_01_5_8PSK}
\end{figure}
\begin{figure}[t]
	\centering
	\includegraphics[width=0.9\linewidth]{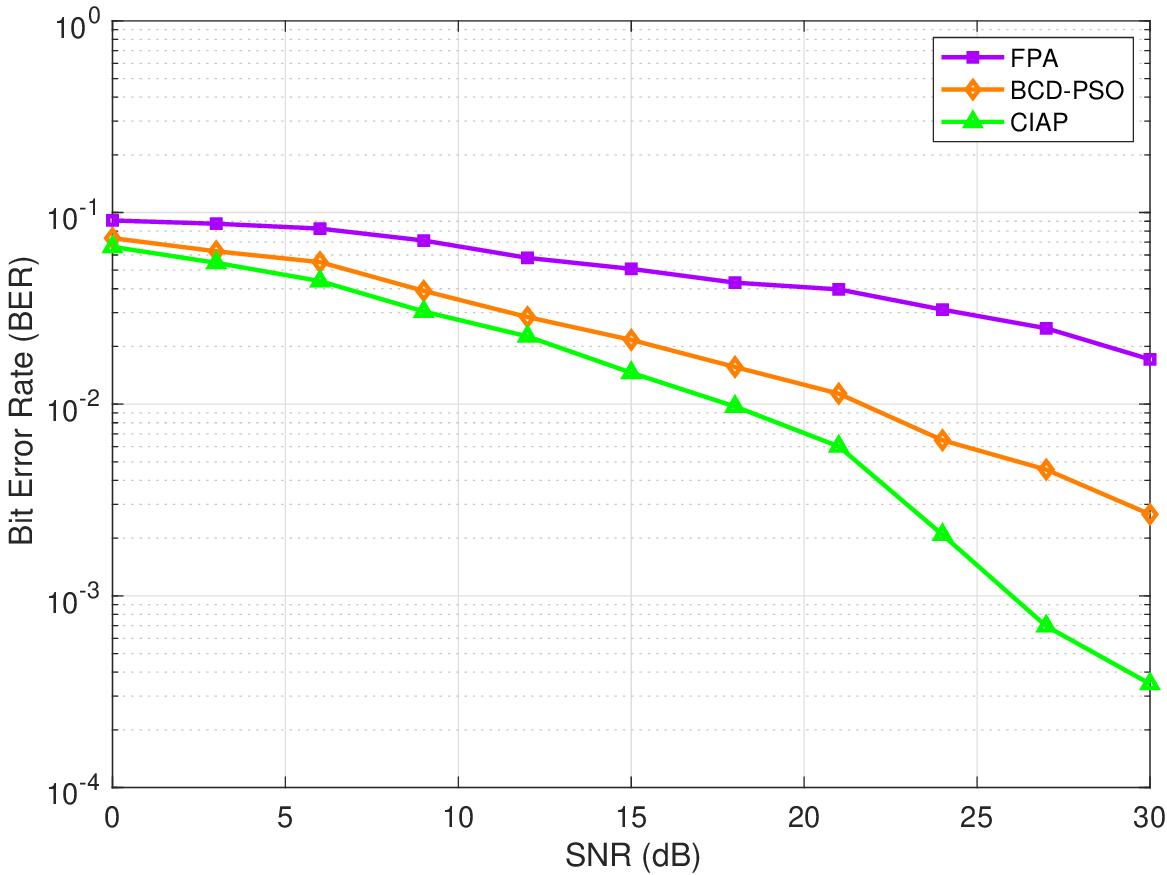}
	\caption{Average BER performance versus SNR; 16PSK}
	\label{fig:10_10_01_5_16PSK}
\end{figure}
\begin{figure}[t]
	\centering
	\includegraphics[width=0.9\linewidth]{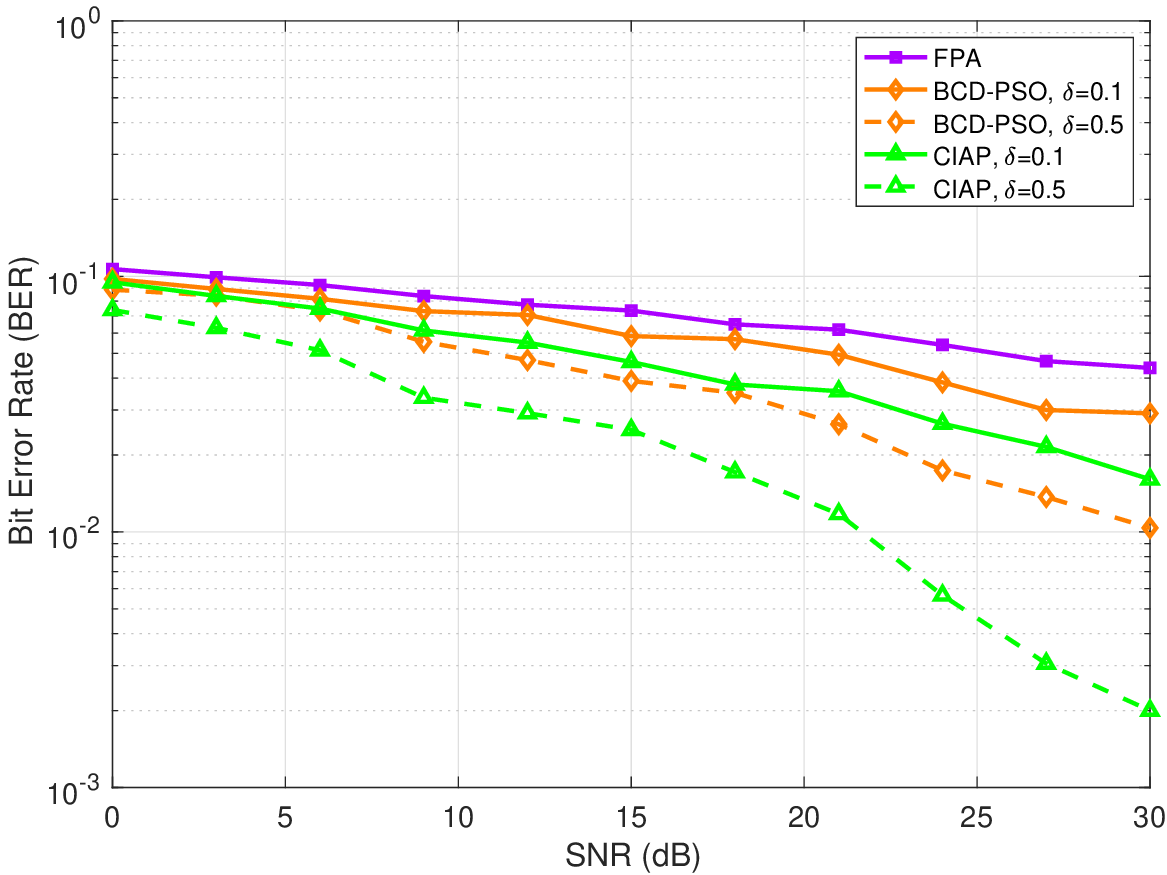}
	\caption{Average BER performance for different $\delta$}
	\label{fig:delta}
\end{figure}
\begin{figure}[ht]
	\centering
	\includegraphics[width=0.9\linewidth]{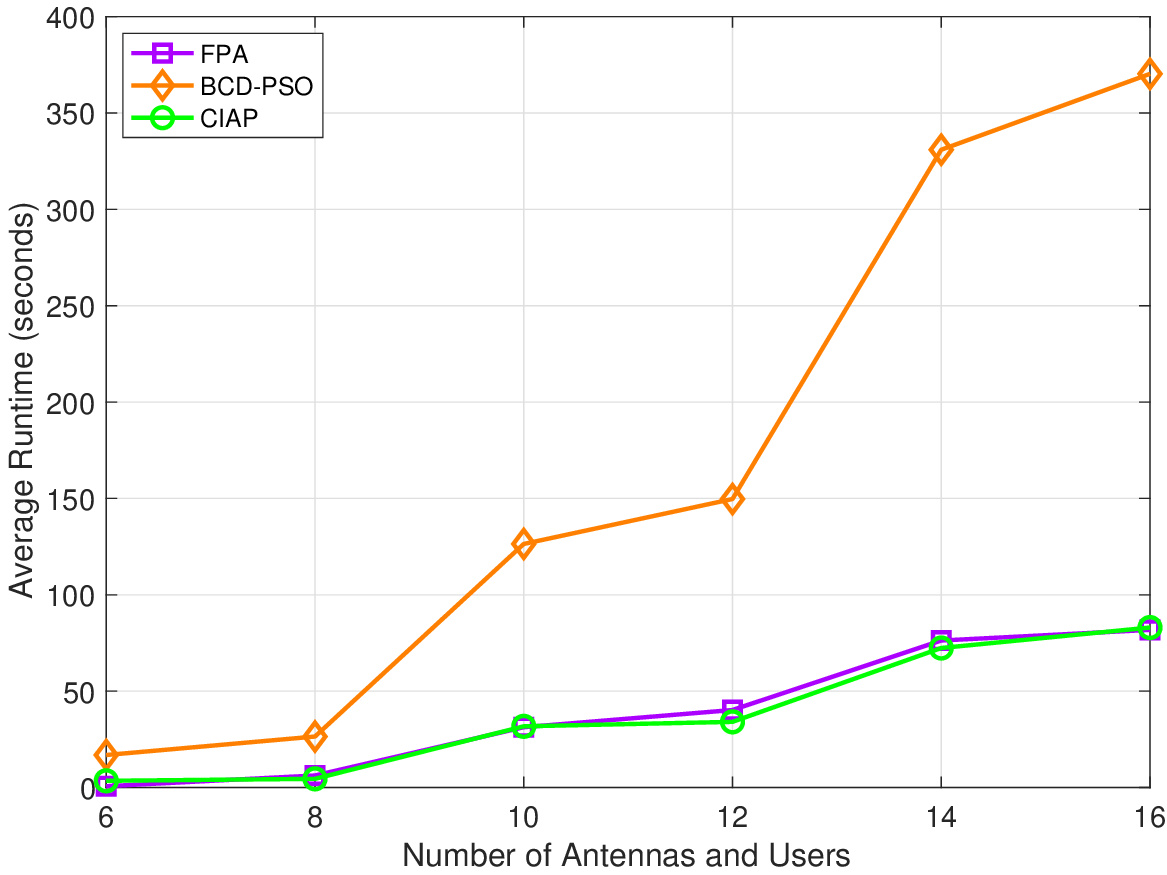}
	\caption{Average runtime in different problem sizes}
	\label{fig:time}
\end{figure}

Figs.~\ref{fig:10_10_01_5_4PSK}-\ref{fig:10_10_01_5_16PSK} illustrate the BER performance 
of the proposed CIAP, BCD-PSO, and FPA   versus the SNR under different MPSK modulations. The number of users and antennas is set to $K=N=8$. 
It is seen that CIAP and BCD-PSO, which optimizes the antenna positions, achieve enhanced performance over the FPA baseline under all different modulation schemes. 
In particular,  comparing the two fluid antenna methods, the proposed CIAP
algorithm achieves better BER performance than BCD-PSO, particularly at higher SNR values. Also, the BER performance degrades for all methods as the modulation order $M$ increases.
However, it is seen that when the modulation order $M$ grows, the performance gain between our proposed CIAP and the BCD-PSO becomes more significant. 
Later, we will also show that our proposed CIAP is much more computationally efficient than BCD-PSO.

Next, we investigate the impact of the size of the antenna position region. 
We fix the total range $D$ unchanged and alter the inter antenna spacing controlled by the parameter $\delta$. 
We simulate the algorithms for 
$N=10$, $K=10$ with 16PSK modulation, comparing the BER performance for $\delta=0.1$ and $\delta=0.5$. 
As shown in Fig. \ref{fig:delta}, increasing the flexibility of antenna positions from $\delta=0.1$ to $\delta=0.5$ leads to a further improvement in BER performance. 
Compared with BCD-PSO, our algorithm exhibits superior BER performance when dealing with a larger adjustment range.

Finally, we compare the average runtime per channel realization for the three algorithms. 
We fix the modulation to 8PSK and vary the number of antennas and users simultaneously ($N=K$) from $6$ to $16$. 
Fig. \ref{fig:time} shows that the runtime of all algorithms increases with the problem size. 
It is seen that for both CIAP and BCD-PSO, optimizing the antenna positions requires additional computational overhead compared to the fixed antenna case FPA. 
However, it is seen that the increased computational burden of our CIAP is marginal, and is much less than that of BCD-PSO.

\section{Conclusion}

This paper has studied a multiuser MISO system employing structured 
fluid antennas under MPSK signaling, with emphasis on the joint optimization of CI precoding and FA positions, leveraging the benefits of two recent advanced physical-layer techniques. 
The resulting problem shows a nonconvex and nonsmooth optimization challenge.
We tackled this challenge by utilizing a smoothing technique, and an efficient 
BCD algorithm combining APG for precoding and PGD for antenna positioning. 
By numerical simulations, we showed the enhanced BER performance compared to traditional fixed antenna arrays. 
Also, the proposed algorithm outperforms the state-of-the-art FA design using PSO. 
It is also worth noting that the performance gain of our proposed algorithm is achieved with a lightweight computational complexity. 

\bibliographystyle{ieeetr}
\bibliography{conference_101719}
\end{document}

%% file: sym_marco.tex
\newcommand\bx{\ensuremath{{\bm x}}}

\newcommand\bz{\ensuremath{{\bm z}}}

\newcommand\bv{\ensuremath{{\bm v}}}

%% file: conference_101719.bbl
\begin{thebibliography}{10}

\bibitem{elmossallamy2020reconfigurable}
M.~A. ElMossallamy, H.~Zhang, L.~Song, K.~G. Seddik, Z.~Han, and G.~Y. Li, ``{Reconfigurable intelligent surfaces for wireless communications: Principles, challenges, and opportunities},'' {\em IEEE Trans. Cognit. Commun. Netw.}, vol.~6, no.~3, pp.~990--1002, 2020.

\bibitem{shao2020minimum}
M.~Shao, Q.~Li, and W.-K. Ma, ``Minimum symbol-error probability symbol-level precoding with intelligent reflecting surface,'' {\em IEEE Wireless Commun. Lett.}, vol.~9, no.~10, pp.~1601--1605, 2020.

\bibitem{pan2022overview}
C.~Pan, G.~Zhou, K.~Zhi, S.~Hong, T.~Wu, Y.~Pan, H.~Ren, M.~Di~Renzo, A.~L. Swindlehurst, R.~Zhang, {\em et~al.}, ``{An overview of signal processing techniques for RIS/IRS-aided wireless systems},'' {\em IEEE J. Sel. Topics Signal Process.}, vol.~16, no.~5, pp.~883--917, 2022.

\bibitem{hao2024fluid}
T.~Hao, C.~Shi, Y.~Guo, B.~Xia, and F.~Yang, ``Fluid-antenna enhanced integrated sensing and communication: Joint antenna positioning and beamforming design,'' in {\em Proc. IEEE/CIC ICCC 2024, \textnormal{2024}}, pp.~956--961.

\bibitem{wong2020fluid}
K.-K. Wong, A.~Shojaeifard, K.-F. Tong, and Y.~Zhang, ``Fluid antenna systems,'' {\em IEEE Trans. Wireless Commun.}, vol.~20, no.~3, pp.~1950--1962, 2020.

\bibitem{new2024tutorial}
W.~K. New, K.-K. Wong, H.~Xu, C.~Wang, F.~R. Ghadi, J.~Zhang, J.~Rao, R.~Murch, P.~Ram{\'\i}rez-Espinosa, D.~Morales-Jimenez, {\em et~al.}, ``{A tutorial on fluid antenna system for 6G networks: Encompassing communication theory, optimization methods and hardware designs},'' {\em IEEE Commun. Surveys Tuts.}, 2024.

\bibitem{wong2023fluid}
K.-K. Wong, W.~K. New, X.~Hao, K.-F. Tong, and C.-B. Chae, ``{Fluid antenna system—part I: Preliminaries},'' {\em IEEE Commun. Lett.}, vol.~27, no.~8, pp.~1919--1923, 2023.

\bibitem{cheng2024sum}
Z.~Cheng, N.~Li, J.~Zhu, X.~She, C.~Ouyang, and P.~Chen, ``Sum-rate maximization for fluid antenna enabled multiuser communications,'' {\em IEEE Commun. Lett.}, vol.~28, no.~5, pp.~1206--1210, 2024.

\bibitem{zhou2024fluid}
L.~Zhou, J.~Yao, M.~Jin, T.~Wu, and K.-K. Wong, ``{Fluid antenna-assisted ISAC systems},'' {\em IEEE Wireless Commun. Lett.}, vol.~13, no.~12, pp.~3533--3537, 2024.

\bibitem{zhang2024efficient}
Q.~Zhang, M.~Shao, T.~Zhang, G.~Chen, J.~Liu, and P.~Ching, ``{An efficient sum-rate maximization algorithm for fluid antenna-assisted ISAC system},'' {\em IEEE Commun. Lett.}, vol.~29, no.~1, pp.~200--204, 2024.

\bibitem{chen2023energy}
Y.~Chen, S.~Li, Y.~Hou, and X.~Tao, ``Energy-efficiency optimization for slow fluid antenna multiple access using mean-field game,'' {\em IEEE Wireless Commun. Lett.}, vol.~13, no.~4, pp.~915--918, 2023.

\bibitem{zhu2024historical}
L.~Zhu and K.-K. Wong, ``Historical review of fluid antenna and movable antenna,'' {\em \textnormal{2024}, arXiv:2401.02362}.

\bibitem{kuang2024movable}
Z.~Kuang, W.~Liu, C.~Wang, Z.~Jin, J.~Ren, X.~Zhang, and Y.~Shen, ``{Movable-antenna array empowered ISAC systems for low-altitude economy},'' in {\em Proc. IEEE/CIC Int.Conf. Commun. China (ICCC Workshops), \textnormal{2024}}, pp.~776--781.

\bibitem{wang2024fluid}
C.~Wang, G.~Li, H.~Zhang, K.-K. Wong, Z.~Li, D.~W.~K. Ng, and C.-B. Chae, ``{Fluid antenna system liberating multiuser MIMO for ISAC via deep reinforcement learning},'' {\em IEEE Trans. Wireless Commun.}, vol.~23, no.~9, pp.~10879--10894, 2024.

\bibitem{spencer2004zero}
Q.~H. Spencer, A.~L. Swindlehurst, and M.~Haardt, ``{Zero-forcing methods for downlink spatial multiplexing in multiuser MIMO channels},'' {\em IEEE Trans. Signal Process.}, vol.~52, no.~2, pp.~461--471, 2004.

\bibitem{zhao2023rethinking}
X.~Zhao, S.~Lu, Q.~Shi, and Z.-Q. Luo, ``{Rethinking WMMSE: Can its complexity scale linearly with the number of BS antennas?},'' {\em IEEE Trans. Signal Process.}, vol.~71, pp.~433--446, 2023.

\bibitem{masouros2009dynamic}
C.~Masouros and E.~Alsusa, ``{Dynamic linear precoding for the exploitation of known interference in MIMO broadcast systems},'' {\em IEEE Trans. Wireless Commun.}, vol.~8, no.~3, pp.~1396--1404, 2009.

\bibitem{jacobsson2017quantized}
S.~Jacobsson, G.~Durisi, M.~Coldrey, T.~Goldstein, and C.~Studer, ``{Quantized precoding for massive MU-MIMO},'' {\em IEEE Trans. Commun.}, vol.~65, no.~11, pp.~4670--4684, 2017.

\bibitem{liu2022symbol}
Y.~Liu, M.~Shao, W.-K. Ma, and Q.~Li, ``Symbol-level precoding through the lens of zero forcing and vector perturbation,'' {\em IEEE Trans. Signal Process.}, vol.~70, pp.~1687--1703, 2022.

\bibitem{shao2019framework}
M.~Shao, Q.~Li, W.-K. Ma, and A.~M.-C. So, ``{A framework for one-bit and constant-envelope precoding over multiuser massive MISO channels},'' {\em IEEE Trans. Signal Process.}, vol.~67, no.~20, pp.~5309--5324, 2019.

\bibitem{sohrabi2018one}
F.~Sohrabi, Y.-F. Liu, and W.~Yu, ``{One-bit precoding and constellation range design for massive MIMO with QAM signaling},'' {\em IEEE J. Sel. Topics Signal Process.}, vol.~12, no.~3, pp.~557--570, 2018.

\bibitem{baghdady2002directional}
E.~J. Baghdady, ``Directional signal modulation by means of switched spaced antennas,'' {\em IEEE Trans. Commun.}, vol.~38, no.~4, pp.~399--403, 2002.

\bibitem{kalantari2016directional}
A.~Kalantari, M.~Soltanalian, S.~Maleki, S.~Chatzinotas, and B.~Ottersten, ``{Directional modulation via symbol-level precoding: A way to enhance security},'' {\em IEEE J. Sel. Topics Signal Process.}, vol.~10, no.~8, pp.~1478--1493, 2016.

\bibitem{li2018interference}
A.~Li and C.~Masouros, ``{Interference exploitation precoding made practical: Optimal closed-form solutions for PSK modulations},'' {\em IEEE Trans. Wireless Commun.}, vol.~17, no.~11, pp.~7661--7676, 2018.

\bibitem{masouros2015exploiting}
C.~Masouros and G.~Zheng, ``Exploiting known interference as green signal power for downlink beamforming optimization,'' {\em IEEE Trans. Signal Process.}, vol.~63, no.~14, pp.~3628--3640, 2015.

\bibitem{chen2023joint}
X.~Chen, B.~Feng, Y.~Wu, D.~W.~K. Ng, and R.~Schober, ``{Joint beamforming and antenna movement design for moveable antenna systems based on statistical CSI},'' in {\em Proc. IEEE Global Commun. Conf. (Globecom), Kuala Lumpur, Malaysia, \textnormal{2023}}, pp.~4387--4392.

\bibitem{shao2018multiuser}
M.~Shao, Q.~Li, Y.~Liu, and W.-K. Ma, ``{Multiuser one-bit massive MIMO precoding under MPSK signaling},'' in {\em Proc. IEEE Global Conf. Signal Inf. Process., \textnormal{2018}}, pp.~833--837.

\bibitem{wu2023diversity}
Z.~Wu, J.~Wu, W.-K. Chen, and Y.-F. Liu, ``Diversity order analysis for quantized constant envelope transmission,'' {\em IEEE Open J. Signal Process.}, vol.~4, pp.~21--30, 2023.

\bibitem{nesterov2005smooth}
Y.~Nesterov, ``Smooth minimization of non-smooth functions,'' {\em Math. Program.}, vol.~103, no.~1, pp.~127--152, 2005.

\bibitem{boyd2004convex}
S.~Boyd, {\em Convex optimization}.
\newblock Cambridge, U.K.: Cambridge Univ. Press, 2004.

\bibitem{zhu2023modeling}
L.~Zhu, W.~Ma, and R.~Zhang, ``Modeling and performance analysis for movable antenna enabled wireless communications,'' {\em IEEE Trans. Wireless Commun.}, vol.~23, no.~6, pp.~6234--6250, 2023.

\end{thebibliography}
